\documentclass{PHYEAUTH}
\usepackage{graphicx}
\usepackage{amsmath}
\usepackage{amssymb}

\begin{document}

\begin{frontmatter}

\title{Vortex formation in quantum dots in high magnetic fields}

\author{H. Saarikoski, A. Harju, M.~J. Puska, and R.~M. Nieminen}

\address{Laboratory of Physics, Helsinki University of Technology,
P.O. Box 1100, FIN-02015 HUT, Finland}

\begin{abstract}
We study electronic structures of two-dimensional quantum dots
in high magnetic fields using the density-functional theory (DFT) and the
exact diagonalization (ED). With increasing magnetic field, beyond the
formation of the totally spin-polarized maximum density droplet (MDD) state,
the DFT gives the ground-state total angular momentum as a continuous function
with well-defined plateaus. The plateaus agree well with the magic angular
momenta of the ED calculation. 
By constructing a conditional wave function from the Kohn-Sham states
we show that vortices enter the quantum dot one-by-one at the transition
to the state with the adjacent magic angular momentum. Vortices are also
observed outside the high-density region of the quantum dot.
These findings are compared to the ED results and
we report a significant agreement.
We study also interpretations and limitations of the density functional
approach in these calculations.
\end{abstract}

\end{frontmatter}

Quasi-two-dimensional (2D) quantum dots fabricated into semiconductor
heterostructures are expected to be basic components in the
future nanoelectronics \cite{jacak}. Moreover, they provide a
unique playground in physics with a rich diversity of quantum effects
among interacting electrons \cite{reimann}.
The reason is especially the strong influence of magnetic fields
on the electron states. Already for fields attainable in
the laboratory environment the interaction with electrons
is of the same order as the electron-electron interactions.
The understanding of experimental findings in quantum dots
and the prediction of new phenomena
have inspired a huge amount of theoretical and computational
work.

Recent developments in the electronic structure theory of 2D
quantum dots in magnetic fields have shown a rich variety of phenomena
related to vortices~\cite{saarikoski,tavernier,pfannkuche,toreblad}.
These vortices correspond to magnetic field quanta
with rotating currents of charge around them and a zero in the wave function
at the vortex centra. By going around one vortex the wave function
gains a phase change of $2\pi n$, where $n$ is the winding number. 
In the fractional quantum Hall effect (FQHE) the state of the matter
is described by Laughlin wave functions which attach
vortex zeros at each electron.
Electronic structure calculations of charge droplets in quantum dots
have predicted the appearance of additional free vortices
which cluster near electrons~\cite{saarikoski,tavernier}.

In this paper we perform a detailed study of vortex formation
in a circulary symmetric 2D quantum dots in a parabolic external confinement
$V_{{\rm c}}({\bf r}) = {1 \over 2} m^* \omega_0^2 {\bf r}^2$.
The model the system by an effective-mass Hamiltonian
\begin{equation}
H=\left(\sum^N_{i=1} \frac{(-i\hbar \nabla_i+e {\bf A} )^2}{2 m^*}
+V_{{\rm c}}({\bf r_i})\right) + \frac{e^2}{4\pi \epsilon} \sum_{i<j}
\frac{1}{r_{ij}} \ ,
\label{hamiltonian}
\end{equation}
where $N$ is the number of electrons in the dot, ${\bf A}$ is the
vector potential of the external perpendicular magnetic field $B$,
$m^*$ the effective electron mass, and $\epsilon$ is the dielectric
constant of the medium.  We use the material parameters of
GaAs, $m^*/m_e=0.067$ and $\epsilon/\epsilon_0= 12.4$.
In high magnetic fields the quantum dot is spin-polarized due to the
Zeeman effect and the maximum density droplet state (MDD)
is formed~\cite{macdonald}.
This state is a finite-size precursor of the $\nu=1$ quantum Hall state.
In parabolic external confining potential the state
is formed by the lowest-Landau-level (LLL)
orbitals with angular momenta $l=0,-1,\dots,-N+1$, and the total
angular momentum $L$ equals to $-N(N-1)/2$. For a stronger $B$, ED has
shown ground states to occur only at certain ``magic'' $L$ values, and
$L$ shows a stepwise structure as a function of
$B$~\cite{maksym,wojs,seki96JPSJ}.

We solve for the ground state electronic structure using the
spin-density-functional theory (SDFT),
the current-density-functional theory (CDFT) \cite{vignalerasolt},
and the exact diagonalization method (ED).
The density-functional approaches are implemented in the real-space
without symmetry restrictions \cite{saarikoski1}.
The main results were calculated using $128 \times128$ grid points but
similar results were obtained using larger grids up to $256 \times 256$
grid points. High spatial resolution of fine grids is necessary
to accurately describe systems with several vortices inside the quantum dot.
The exchange-correlation effects are taken into account using
local approximations for the spin densities and
the vorticity in the 2D electron gas~\cite{attaccalite,saarikoski2}.

The Hamiltonian of the system is rotationally symmetric
and therefore the particle density, defined as $|\Psi({\bf r})|^2$,
should also be rotationally symmetric.
Calculations, however, show that
the particle density in the density-functional formalism is
{\em not} necessarily symmetric.
In this work we adopt the approach that even if broken symmetry
solutions cannot be identified as the true particle density we
may use them to find out correlations in the system. Otherwise,
the correlations may remain hidden in the one-particle picture of the density
functional approach.
On the other hand one, should be cautious with the interpretations since
it has been shown that the symmetry breaking solutions
may be wrong mixtures of the eigenstates of the system and
therefore the interpretation of the solution may be unphysical \cite{ari}.
We therefore emphasize that solutions of the DFT
should be used as guidelines and approximations and the final proofs
should be done using exact many-body techniques.

We examine the 6-electron quantum dot $(N=6)$.
For an easy comparison with the low magnetic field results in
Ref. \cite{saarikoski1} we set the confinement strength to
$\omega_0=5\; {\rm meV}$.
The maximum density droplet (MDD) state is characterized
by a relatively constant electron density and currents concentrated at the
edge of the dot. 
The transition to the MDD state occurs at
$4.8\;{\rm T}$.

\begin{figure}
\includegraphics[width=.96\columnwidth]{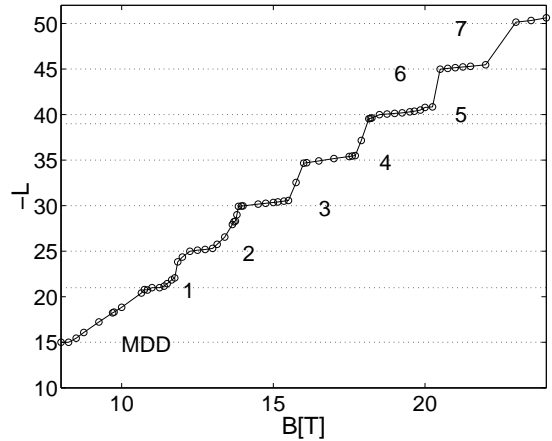}
\caption{Angular momentum $L$ of the six-electron QD from SDFT (open circles).
The plateaus are characterized by vortex-holes in the electron density.
The number next to a plateau gives the number of vortices {\it inside} the QD.
The horizontal lines correspond to ground state magic $L$ values from ED.
Solutions between the plateaus of nearly constant $L$ are charge-density-wave
like solutions with fractional $L$.
}
\label{fig:lz}
\end{figure}

The total angular momentum in the SDFT in the beyond-MDD domain is shown
in Fig. \ref{fig:lz}. The curve has plateaus of
nearly constant $L$. The plateau regions are characterized by
electron densities with vortex depressions inside the quantum dot and
outside the dot in the low density region.
The number of vortices inside the quantum dot increases by one
between adjacent plateaus.
The plateaus correspond well to the
magic angular momenta from the ED calculations.
In the ED four-vortex configuration has the pentagon or hexagon symmetry which
correspond to angular momenta $L=-35$ and $L=-39$, respectively.
In the DFT solutions, however, only the $L=-35$ plateau is visible.

\begin{figure}
\includegraphics[width=0.48\columnwidth,height=0.48\columnwidth]{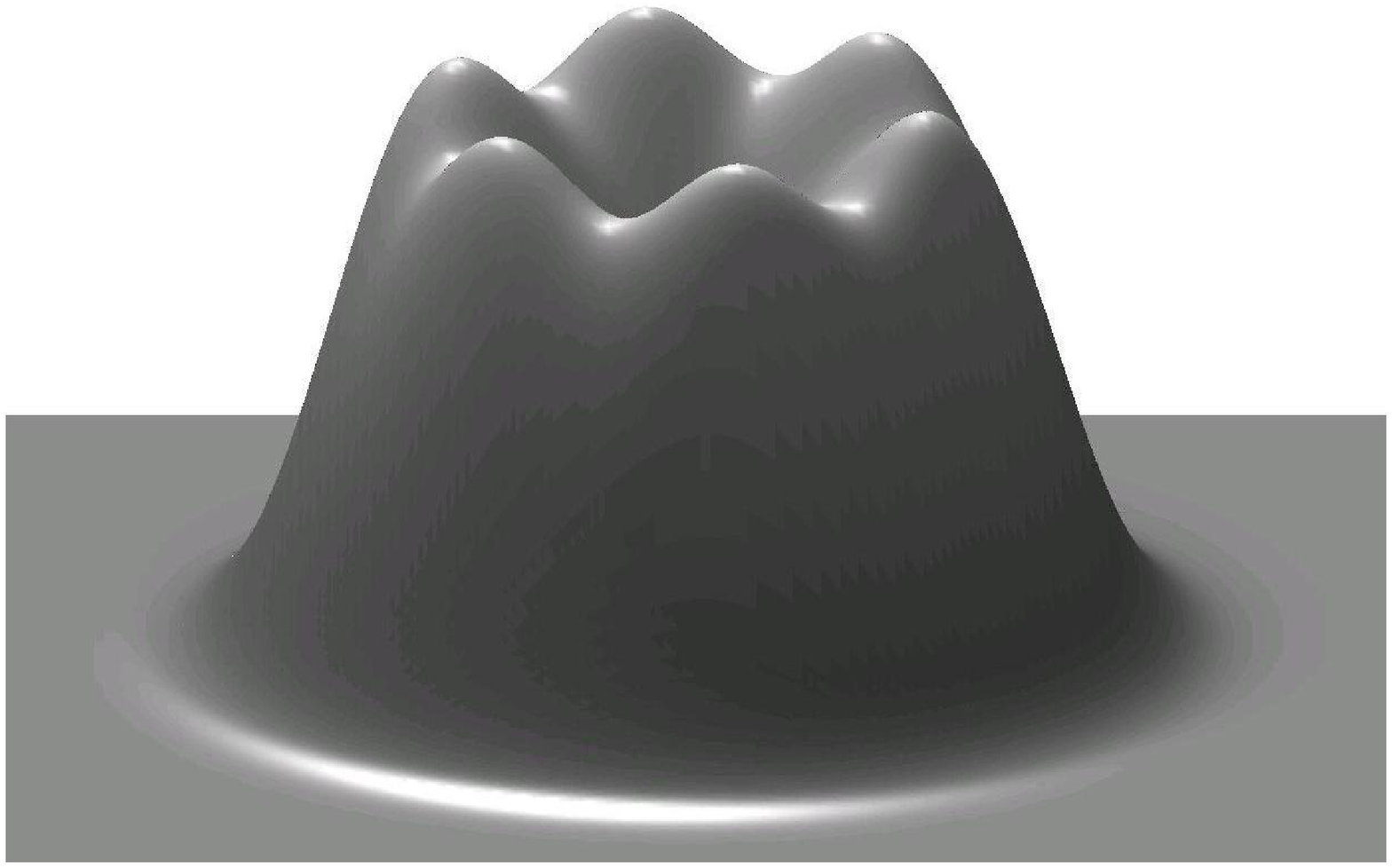}
\includegraphics[width=0.48\columnwidth,height=0.48\columnwidth]{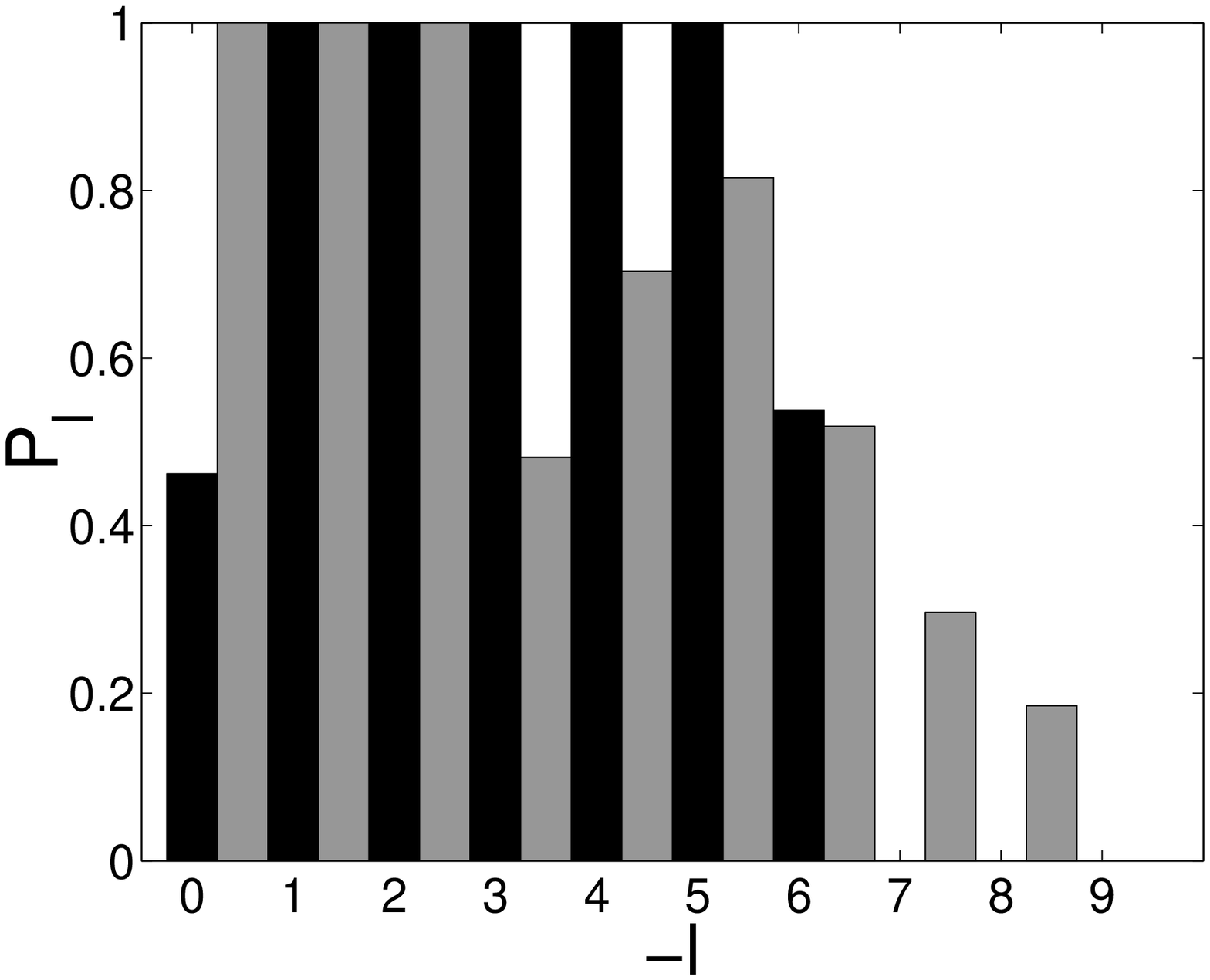}
\caption{Charge-density-wave (CDW) solution in a six-electron quantum dot
at 9.7 T. The total angular momentum $L\approx-18.230$.
Left panel: the electron density. Right panel: the projections of the angular
momentum to Fock-Darwin states for the
CDW solution (black bars) are compared to the $L=-18$ solution
from the exact diagonalization method (grey bars).
}
\label{fig:cdw}
\end{figure}

In the density-functional solutions vortices appear as static holes which break
the rotational symmetry of the particle density.
The wave function must vanish at the vortex position in order to keep the
kinetic energy finite.
In the beyond MDD-domain there are also charge-density-wave (CDW)-like
solutions
with fractional $L$ values in the transition regions between the
adjacent magic angular
momentum values. For example, solutions from $B\simeq 8.3\;{\rm T}$ to
$B\simeq 11\;{\rm T}$ have 6 density maxima at the edge of the dot
and a minimum at the dot center (see Fig. \ref{fig:cdw})

We do a detailed analysis of the vortex solutions in the SDFT by
constructing a conditional wave function of the Kohn-Sham states.
We do it iteratively by fixing $N-1$ electrons 
to positions $\{{\bf r}^*_i\}^N_{i=2}$ and then calculate
a determinant of the $N$ occupied Kohn-Sham
states $\phi_1,...,\phi_N$
\begin{equation}
\Phi_{\rm KS}({\bf r})
=
\left|
\begin{array}{cccc}
\phi_1({\bf r}) & \phi_2({\bf r}) & \ldots & \phi_N({\bf r}) \\
\phi_1({\bf r}^*_2) & \phi_2({\bf r}_2^*) & \ldots & \phi_N({\bf r}_2^*) \\
\vdots & \vdots & \ddots & \vdots \\
\phi_1({\bf r}^*_N) & \phi_2({\bf r}_N^*) & \ldots & \phi_N({\bf r}_N^*)
\end{array}
\right|.
\label{determinant}
\end{equation}
The electron positions are moved until the total probability of
the $\Phi_{\rm KS}$ obtains a maximum value.
Strictly speaking $\Phi_{\rm KS}({\bf r})$ constructed in this
way is only an auxiliar function which emulates the true conditional
many-body wave function~\cite{saarikoski} which is unknown in
the density functional approach.

\begin{figure}
\includegraphics[width=.48\columnwidth]{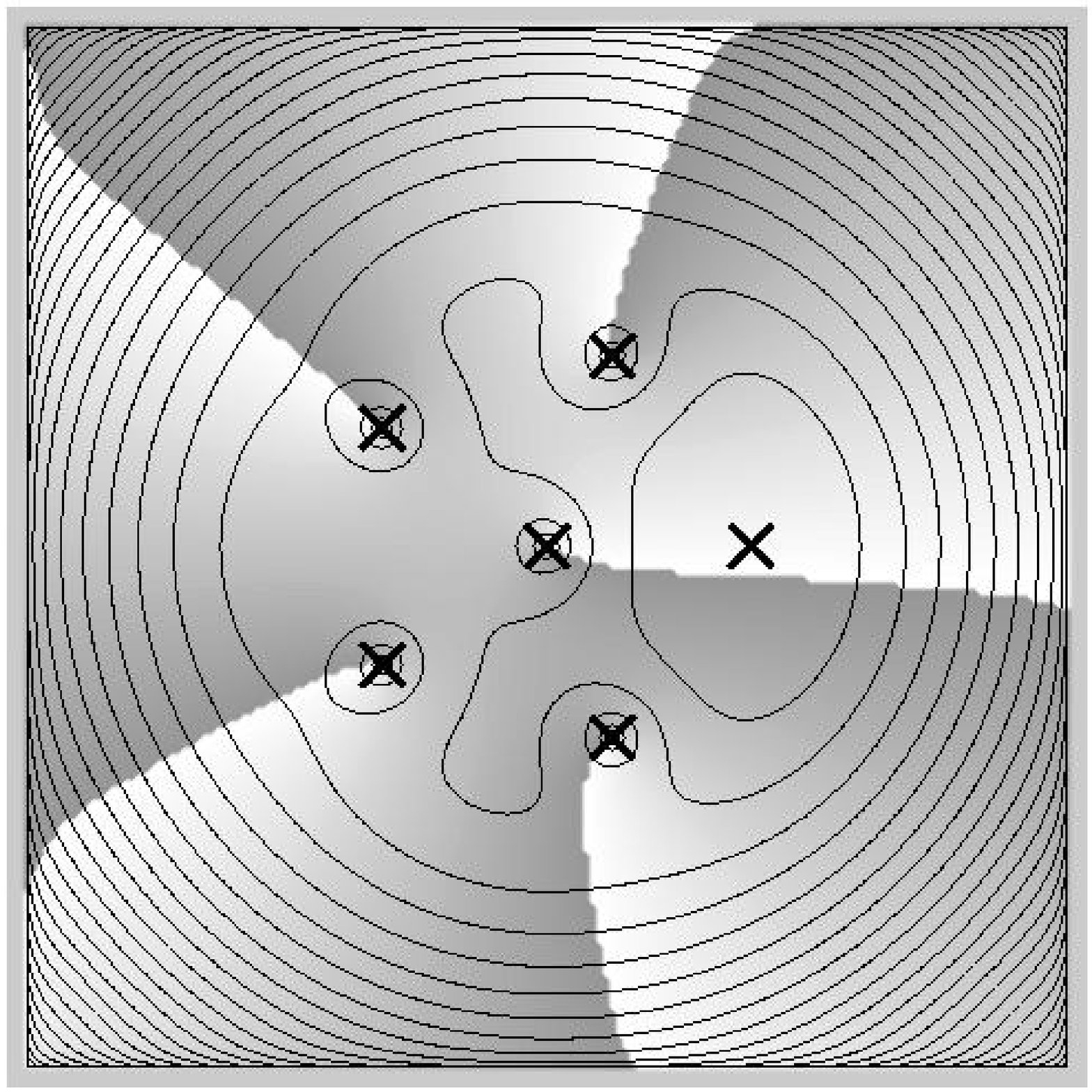}
\includegraphics[width=.48\columnwidth]{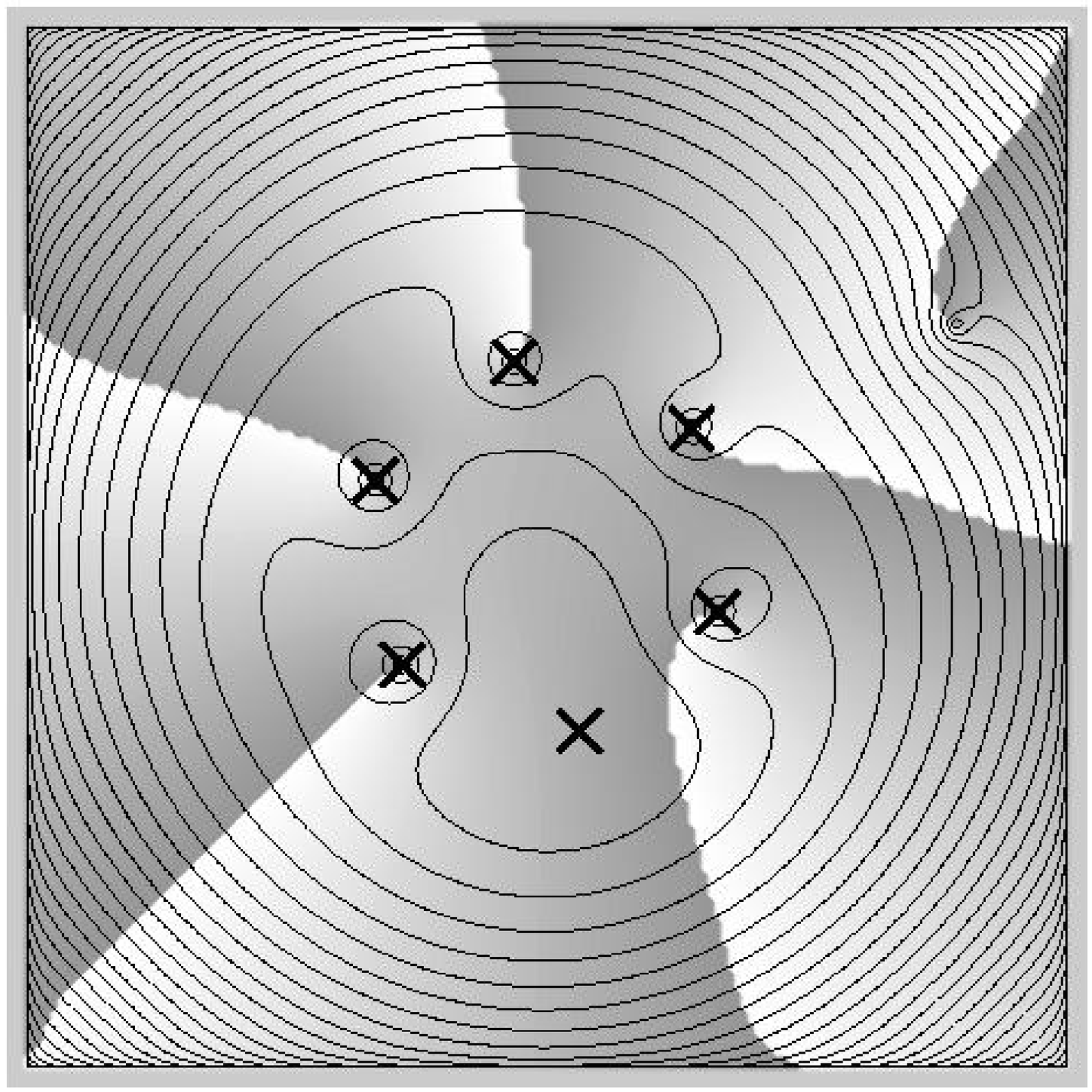}
{\hspace{0.2\columnwidth}}
MDD, $L=-15$
{\hspace{0.2\columnwidth}}
CDW, $L\approx -16.103$
\includegraphics[width=0.48\columnwidth]{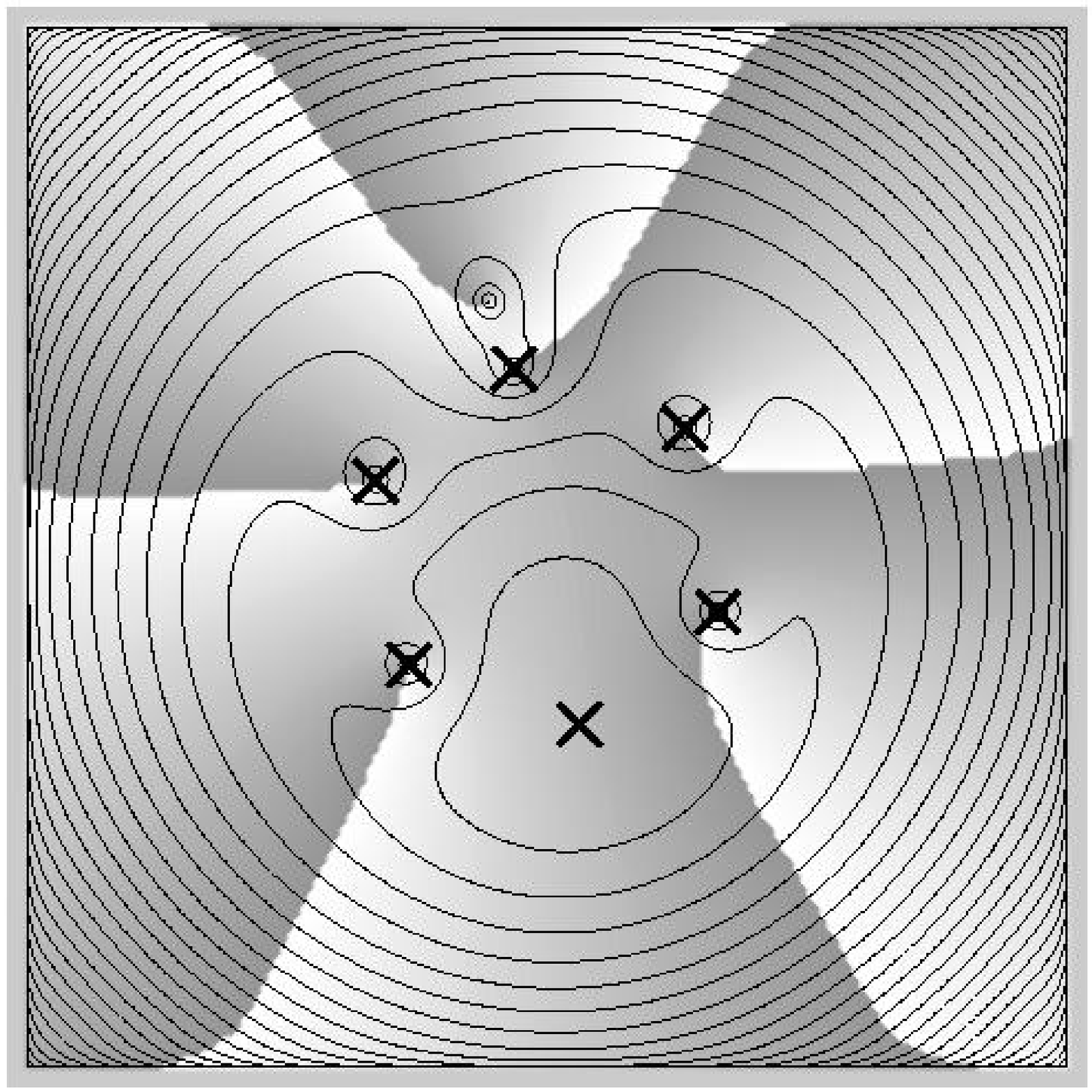}
\includegraphics[width=0.48\columnwidth]{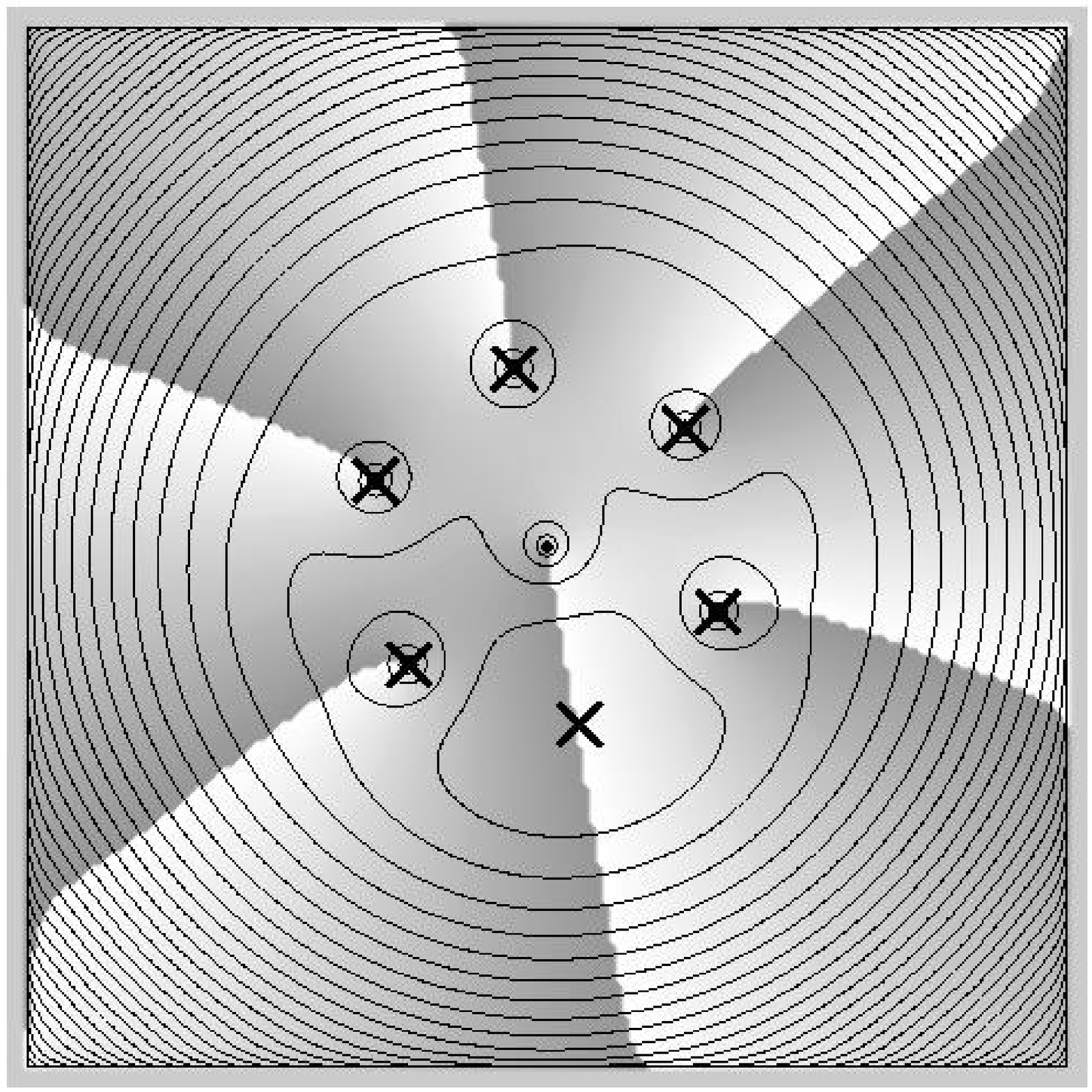}
{\hspace{0.1\columnwidth}}
CDW, $L\approx -17.258$
{\hspace{0.15\columnwidth}}
Single-vortex, $L=-21$
\caption{Conditional wave function from the Kohn-Sham states
$\Phi_{\rm KS}({\bf r})$ for the MDD-state, two charge-density-wave
states, and a single-vortex state.
The contour plots are logarithmic conditional electron densities,
the shadowings show the phases of the wave-function, and the crosses
mark the most probable electron positions.
At the lines where shadowing changes from the darkest grey to the lightest grey
the phase changes by $2\pi$.
}
\label{fig:singlevortex}
\end{figure}

$\Phi_{\rm KS}({\bf r})$ for the MDD-state correctly assigns one vortex
for each electron. The evolution of this state as the magnetic field
is increased is shown in Fig. {\ref{fig:singlevortex}}.
It depicts a vortex hole entering the quantum dot as the magnetic field
is increased. The CDW solutions can be analyzed further by
projecting the Kohn-Sham states on the Fock-Darwin states \cite{reimann}
and comparing the results with those from the exact diagonalization
(Fig. \ref{fig:cdw}). From this data we can conclude that
the charge-density-wave originates from the combinations of the
$L=-15$ and $L=-21$ states.  The mixture of these two states result in
a density with six peaks in a form of a hexagon.
The linear behaviours between plateaus can be
thought to be unphysical since the exact ground state jumps directly
from the MDD to $L=-21$.
However the DFT gives the same qualitative picture than the ED
where the (excited state) solutions between $L=-15$ and $-21$ show
a vortex-hole moving from the outer edge toward the center~\cite{oaknin}.
The other CDW-regions show similar solutions, transport of one
additional vortex-hole to the center of the dot.

\begin{figure}
\includegraphics[width=.48\columnwidth]{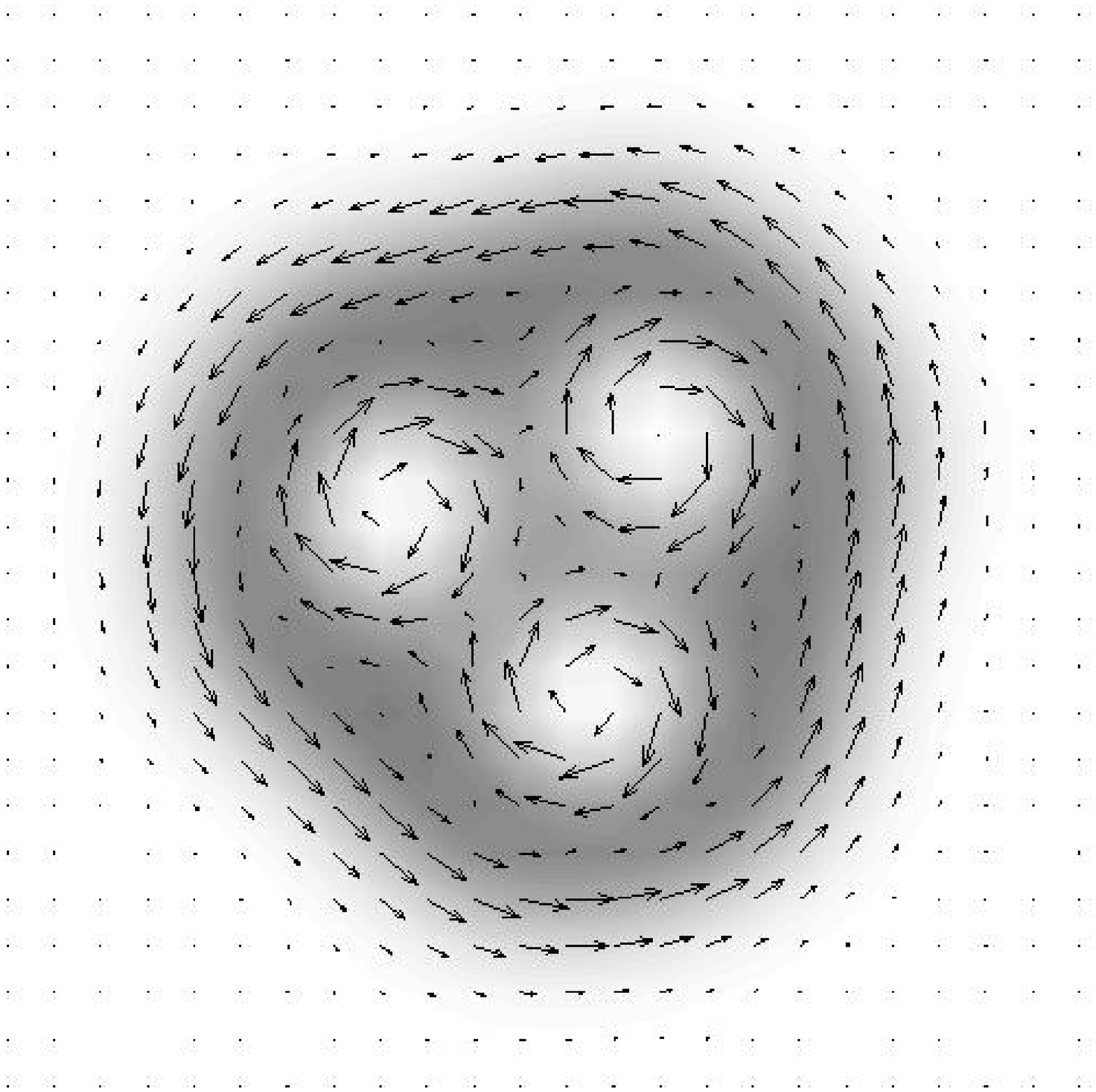}
\includegraphics[width=.48\columnwidth]{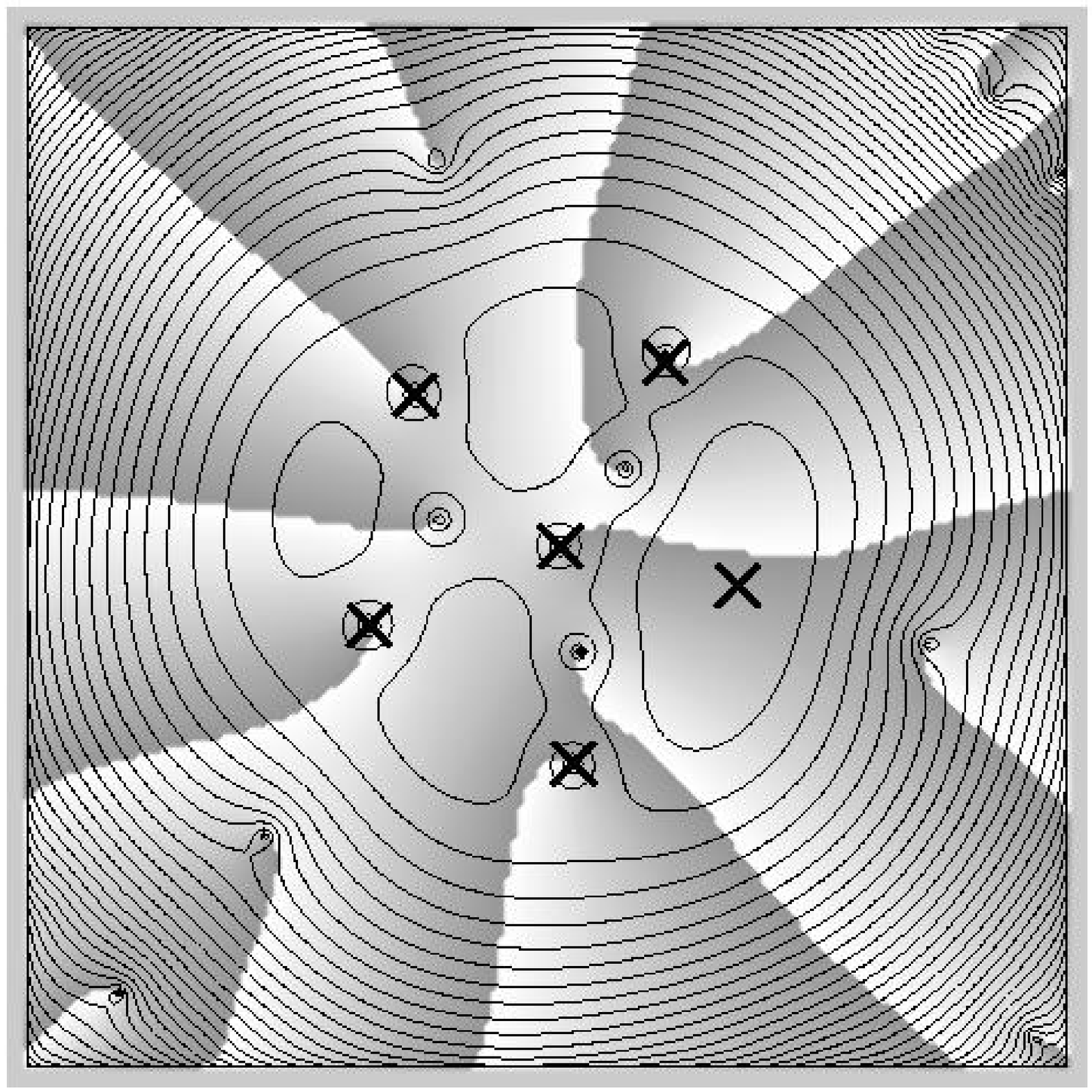}
\caption{Electron density (grey scale) and the physical current density
(vectors) for the CDFT triple-vortex solution at 15.5 T (left).
Conditional wave function from the Kohn-Sham states
$\Phi_{\rm KS}({\bf r})$ for the triple-vortex solution (right).
The pictures have not the same scale.
}
\label{fig:3vrtx}
\end{figure}

\begin{figure}
\includegraphics[width=.48\columnwidth]{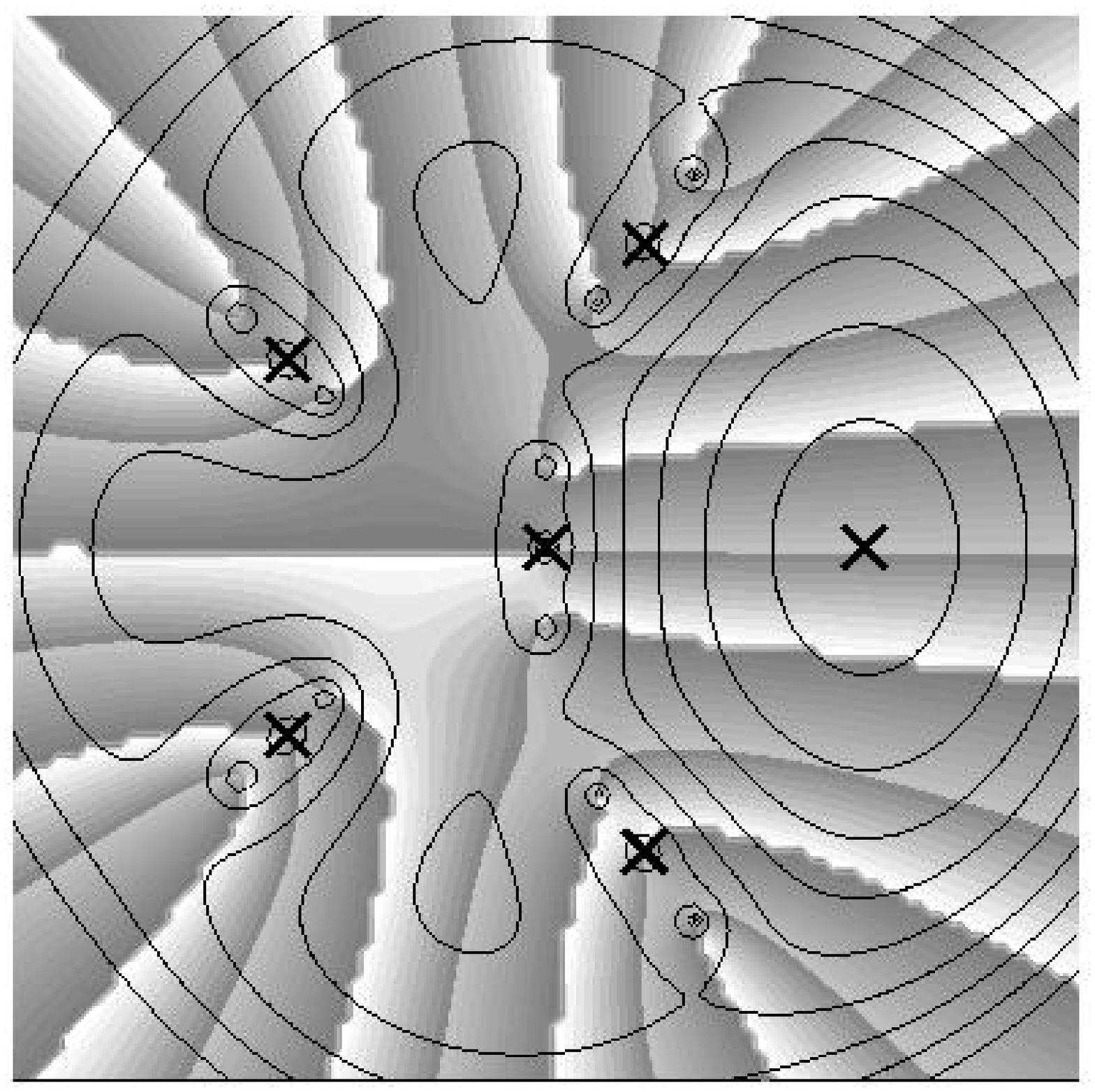}
\includegraphics[width=.48\columnwidth]{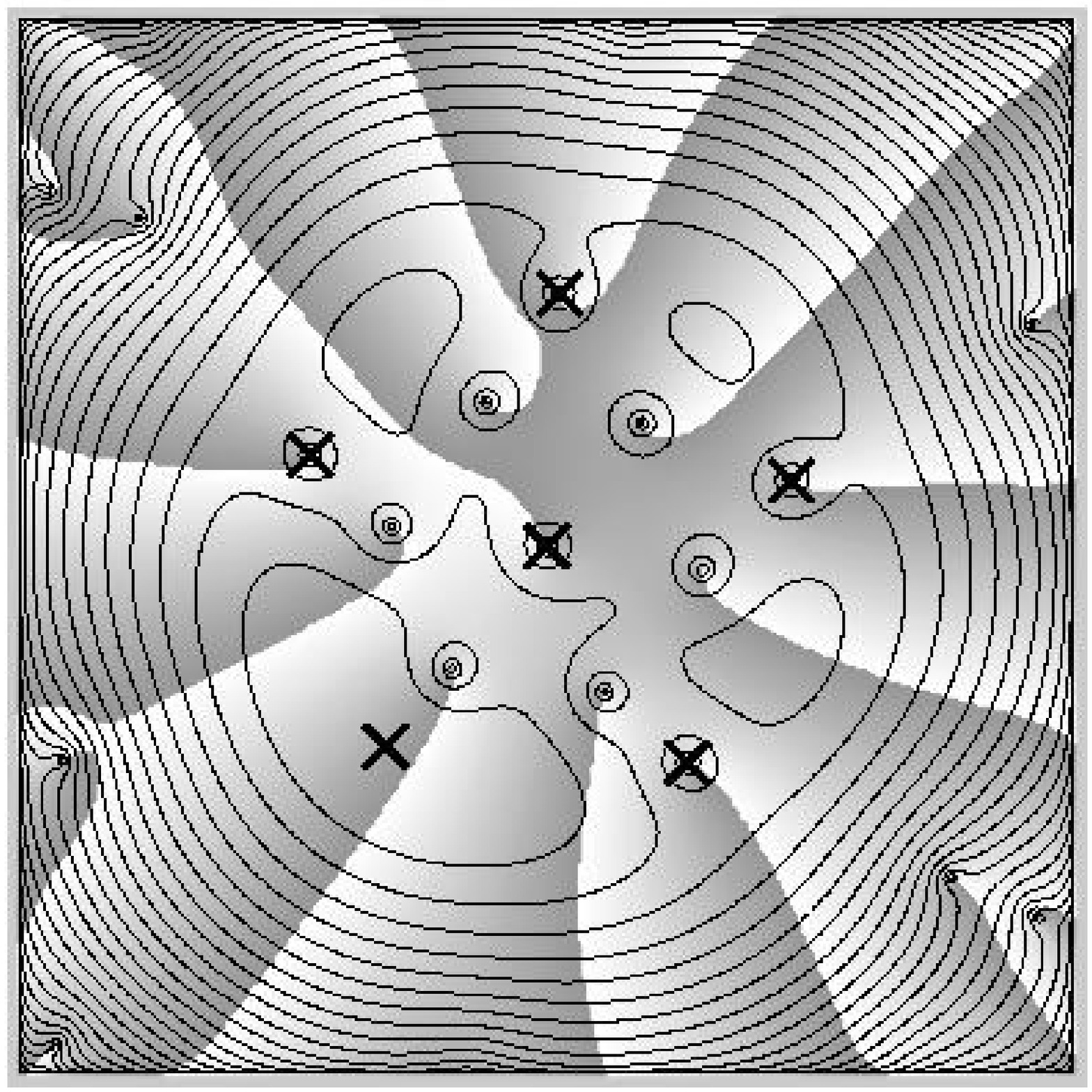}
\caption{Conditional ED wave function of $L=-45$ state (left) and the 
$\Phi_{\rm KS}({\bf r})$ for the $L\simeq -45.129$ state at 21 T (right).
}
\label{fig:6vrtx}
\end{figure}

The triple-vortex solution at $B=15.5\;{\rm T}$ is shown in
Fig. \ref{fig:3vrtx}.
The DFT localizes the vortices in the electron density.
The conditional wave function shows vortices at exactly these positions.
External vortices appear both in the conditional wave function and
in the logarithmic plot of the electron density (not shown).
The comparison with the ED results
for $L=-30$ in Ref. \cite{saarikoski} shows that
the conditional DFT wave function given by Eq. (\ref{determinant})
correctly predicts the existence of three vortices of winding number one
inside the ring of the fixed electrons.
In the ED the vortices start to cluster around the
fixed electrons for larger angular momenta~\cite{saarikoski,tavernier}.
The conditional DFT wavefunction (\ref{determinant}) predicts the right
amount of vortices inside the quantum dot but fails to
predict the clustering of vortices. This can be seen by comparing the ED
and DFT results for the $L=-45$ case in Fig. \ref{fig:6vrtx}.
Therefore the $L=-45$ state can be identified as a finite size
precursor of the $\nu=1/3$ FQHE state only in the ED.

To conclude, we have studied vortex formation in parabolically
confined quantum dots
and found that both the DFT and the ED predict the existence of solutions with
free vortices moving between the electrons. We have analysed 
the structure of these solutions using
conditional wave functions and we
found that the DFT gives qualitatively right results but is unable to predict
the clustering of vortices near the electrons.
We hope that these results will encourage further analysis of vortex
solutions e.g. in systems of large electron number and low
symmetry~\cite{ellipse}.

We thank Dr. Stephanie Reimann for fruitful ideas and discussions.
This work has been supported by Academy of Finland through the Centre
of Excellence Program (2000-2005).

\end{document}